\documentclass[11pt,letterpaper]{article}
\usepackage[utf8]{inputenc} 
\usepackage[T1]{fontenc}    
\usepackage[margin=1in]{geometry}

\usepackage[colorlinks=true,linkcolor=blue,citecolor=blue]{hyperref}
\usepackage{amsfonts}
\usepackage{amsthm}
\usepackage{mathtools}
\usepackage{amssymb}
\usepackage{thmtools}
\usepackage{thm-restate}
\usepackage{cleveref}
\usepackage{float} 
\usepackage{soul}
\usepackage{nicefrac}
\usepackage{url}            
\usepackage{booktabs}       

\usepackage{algorithm}
\usepackage{algpseudocode}

\newtheorem{lemma}{Lemma}

\newtheorem{proposition}{Proposition}
\newtheorem{corollary}{Corollary}

\theoremstyle{remark}
\newtheorem*{remark}{Remark}

\usepackage{xcolor}
\usepackage{bbm}
\allowdisplaybreaks

\title{On the Dissipation of Ideal Hamiltonian Monte Carlo Sampler}
\author{Qijia Jiang\thanks{Lawrence Berkeley National Laboratory, \texttt{qjiang@lbl.gov}.}}  
\date{\today} 

\begin{document}

\maketitle

\begin{abstract}
We report on what seems to be an intriguing connection between variable integration time and partial velocity refreshment of Ideal Hamiltonian Monte Carlo samplers, both of which can be used for reducing the dissipative behavior of the dynamics. More concretely, we show that on quadratic potentials, efficiency can be improved through these means by a $\sqrt{\kappa}$ factor in Wasserstein-2 distance, compared to classical constant integration time, fully refreshed HMC. We additionally explore the benefit of randomized integrators for simulating the Hamiltonian dynamics under higher order regularity conditions.
\end{abstract}

\section{Introduction}
HMC samplers \cite{duane1987hybrid} (and closely related variants \cite{u-turn}) are widely adopted in practice - nice introduction and summary of our current quantitative understanding of them can be found in \cite{vishnoi2021introduction}. Finding its roots in Lattice QCD in the physics community, it has been popularized by \cite{neal2011mcmc} in the computational statistics and machine learning community. Modern implementation of the algorithm has become the choice for the probabilistic programming language Stan \cite{carpenter2017stan}, which has emerged as an integral part of the computational pipeline for any Bayesian statistician. In this work, we study algorithms based on the Hamiltonian dynamics
\begin{equation}
\label{eqn:hmc}    
dx_t = v_t \, dt \quad\quad dv_t = -\nabla f(x_t)\, dt\, ,
\end{equation} 
which is a system of differential equations over the position and velocity variables $(x,v)\in \mathbb{R}^{2d}$. One could also write this as a second order ODE $\ddot{x}(t) = -\nabla f(x(t))$, a familiar equation recognizable from Newtonian mechanics. The fact we refer to it as ideal is because we will assume the Hamiltonian dynamics can be integrated exactly for time period $T$ (where in practice, this is typically done by numerical approximation). The dynamics itself, in general, is not ergodic due to well-documented periodicity arising from energy conservation. However, provided $T$ is not too large and the momentum $v$ is redrawn (in this case from a Gaussian) at appropriate intervals, samplers based on it can simulate long trajectories (i.e., travel far in space), which are viewed as favorable to SDE-based dynamics that injects continual stochasticity exhibiting diffusive behavior. 

Throughout, the mass matrix is assumed to be identity so the stationary distribution over the extended state space is the Boltzmann-Gibbs distribution
\[\pi(x,v)\propto e^{-f(x)-\frac{1}{2}\|v\|^2}\]
and one simply takes the marginal over $x$ to obtain samples from the target $e^{-f}$. Denote $\mathcal{H}(x,v):=f(x)+\frac{1}{2}\|v\|^2$ the (separable) Hamiltonian with potential and kinetic energy, an immediate consequence is that \[\frac{d\mathcal{H}}{dt}=\frac{d\mathcal{H}}{dx}\cdot \frac{dx}{dt} + \frac{d\mathcal{H}}{dv}\cdot \frac{dv}{dt} = 0\] under the dynamics \eqref{eqn:hmc}. To gain some intuition, suppose we start from a position with high potential energy (i.e., low density region), since $\mathcal{H}$ is preserved under the dynamics, if the dynamics is not evolved for too long, a large part of the potential energy will turn into kinetic energy. At this point, the subsequent momentum randomization will, with high probability, decrease the kinetic energy, therefore this occasional randomization induces a dissipative behavior. We will show later that the control of this dissipation (either through traditional velocity refreshment or other means) is crucial to obtain an efficient sampler. As a side remark, it is evident from the exposition above that HMC requires gradient oracle but is oblivious to unknown normalizing constant. 

We mostly focus on sampling from high-dimensional, ill-conditioned quadratic targets in this article. To give a concrete example for its ubiquity, in a class of Bayesian variable selection linear regression model $y=X\beta+\epsilon, \epsilon\sim \mathcal{N}(0,\sigma^2 I)$ using spike and slab prior with a hierarchical structure:
\[z_i\sim \text{Bern}(q), \beta_i|z_i=1 \sim \mathcal{N}(0,\tau_1^2), \beta_i|z_i=0 \sim \mathcal{N}(0,\tau_0^2), \tau_1\gg \tau_0\, ,\]
a typical Gibbs sampler can be set up to alternate between
\[\beta|z\sim \mathcal{N}\left(\beta;(X^\top X+\frac{\sigma^2}{\tau_z^2}I)^{-1}X^{\top}y,\sigma^2(X^\top X+\frac{\sigma^2}{\tau_z^2}I)^{-1}\right)\]
and
\[z_i|\beta_i\sim \text{Bern}\left(\frac{q\mathcal{N}(\beta_i;0,\tau_1^2)}{(1-q)\mathcal{N}(\beta_i;0,\tau_0^2)+q\mathcal{N}(\beta_i;0,\tau_1^2)}\right)\, .\]
The bottleneck of this procedure lies in simulating the first step. Therefore speeding up its implementation is crucial for efficient posterior sampling in this case, which will be important for estimating the unknown sparse $\beta$.

\subsection{Related Work}
We briefly review some previous work here and will mention more in later contexts. For ideal HMC with constant integration time, the work of \cite{chen_vempala2022optimal} gave \emph{tight} rate of $\mathcal{O}(\kappa\log(1/\epsilon))$ in Wasserstein-2 distance for general strongly convex, smooth potentials. Later result of \cite{wang2022accelerating} made the interesting connection to optimization and demonstrated that Chebyshev integration time provably accelerates for quadratic potentials to complexity $\mathcal{O}(\sqrt{\kappa}\log(1/\epsilon))$. We refer the reader to e.g., \cite{bou2021hmc-mixing} for the latest development of discretization analysis for unadjusted HMC. The work of \cite{lu2020explicit} used PDE argument for hypocoercive equation for analyzing the ideal HMC with random integration duration, but it is our understanding that it does not directly render a $W_2$ mixing time guarantee.

\section{Quadratic Potentials}
\label{sec:quadratic}
In this section, we consider $f(x) = \frac{1}{2}x^{\top}\Sigma x$ where $\Sigma$ is a diagonal covariance matrix with $\mu \leq \{\sigma_i\}_{i=1}^d\leq L$ and let $\kappa:=L/\mu > 1$ denote the condition number. We are mostly concerned with the scaling with $\kappa$ in this work. The $x$ marginal of the stationary distribution is denoted as $\pi_x$ and the position output after iteration $K$ is distributed as $x_K\sim\rho_K$.

\subsection{Partial Refreshment}
\label{sec:partial}
In optimization, memory term based on past iterates has long been considered as a hallmark of the acceleration phenomenon. We draw upon this ancient wisdom and examine its effectiveness for HMC sampling. This essentially becomes the Generalized HMC algorithm from \cite{horowitz1991generalized} without the accept/reject correction. 
\begin{algorithm}[H]
\caption{Ideal HMC with friction/damping}\label{alg:momentum}
\begin{algorithmic}
\Require Initial $(x_0,v_0)\in\mathbb{R}^d\times\mathbb{R}^d \sim \rho_0\otimes\mathcal{N}(0,I_d)$ independent \Require Integration time $T>0$, friction $\eta\in[0,1)$, num of iterations $K>1$
\For{$k=1,\cdots,K$}
\State Draw independent $z,z'\sim\mathcal{N}(0,I_d)$
\State $\tilde{v}_{k-1}=\eta v_{k-1}+\sqrt{1-\eta^2} \cdot z$
\State $(x_k,\tilde{v}_k) = \text{HMC}_T(x_{k-1},\tilde{v}_{k-1})$
\State $v_k=\eta \tilde{v}_k+\sqrt{1-\eta^2} \cdot z'$
\EndFor
\State\Return $x_1,\cdots, x_K$
\end{algorithmic}
\end{algorithm}
The first and last step follow an Ornstein-Uhlenbeck process (that doesn't require evaluation of the gradient or affect the position)
\begin{equation}
\label{eqn:damping}    
dv_t = -\gamma v_t\, dt+ \sqrt{2\gamma}\, dB_t
\end{equation}
whose closed-form solution is
\[v_{T/2} = e^{-\gamma {T/2}}v_0+\sqrt{1-e^{-\gamma T}}z\, .\]
For $\gamma = \frac{2}{T}\log(\frac{1}{\eta})$, this draws connection to the under-damped Langevin dynamics (which has explicit friction and dissipation terms in the velocity variable)
\begin{align*}
dx_t = v_t\, dt \quad\quad dv_t = -\nabla f(x_t)\, dt-\gamma v_t \, dt+ \sqrt{2\gamma}\,dB_t
\end{align*}
as it would be a valid symmetrized splitting scheme for its implementation (both part \eqref{eqn:hmc} and \eqref{eqn:damping} will have advanced for time $T$), and is reminiscent of the popular OABAO scheme after discretization. It is clear that in this case, even without adjustment, Algorithm \ref{alg:momentum} admits the right invariant measure sans bias. Through the well trodden path of synchronous coupling, the lemma below shows that it suffices to study the spectral radius of a transition matrix to quantify its convergence rate in $W_2$ metric. 
\begin{lemma}[Coupling of HMC]
\label{lem:harmonic_oscillator}
For quadratic potentials, Algorithm \ref{alg:momentum} updates follow a linear transition $y_+=Ay+BG$, where $G$ is standard normal independent of $y=(x,v)$. Moreover, $W_2(\rho_K,\pi_x)\leq \rho(A^{\top}A)^{K/2}\cdot W_2(\rho_0,\pi_x)$ for $\rho(A^{\top}A) < 1$ the spectral radius of the matrix $A$ specified in the proof.
\end{lemma}
\begin{proof}
Coordinate-wise, using the fact that HMC update for the harmonic oscillator can be written as:
\[x_t(i) = \cos(\sqrt{\sigma_i}t) x_0(i)+\frac{1}{\sqrt{\sigma_i}}\sin(\sqrt{\sigma_i}t)v_0(i)\]
\[v_t(i) = -\sqrt{\sigma_i}\sin(\sqrt{\sigma_i}t)x_0(i)+\cos(\sqrt{\sigma_i}t)v_0(i)\]
which is simply rotating in phase space (i.e., trajectory orbits along constant energy level), we have 
\begin{align*}
\begin{bmatrix}
x_k(i) \\ v_k(i)
\end{bmatrix} &= \begin{bmatrix}
\cos(\sqrt{\sigma_i}T) & \frac{\eta}{\sqrt{\sigma_i}}\sin(\sqrt{\sigma_i}T)\\ -\eta\sqrt{\sigma_i}\sin(\sqrt{\sigma_i}T) & \eta^2\cos(\sqrt{\sigma_i}T)
\end{bmatrix}\begin{bmatrix}
x_{k-1}(i)\\ v_{k-1}(i)
\end{bmatrix}  
+\sqrt{1-\eta^2}\begin{bmatrix}
\frac{1}{\sqrt{\sigma_i}}\sin(\sqrt{\sigma_i}T) & 0\\ \eta\cos(\sqrt{\sigma_i}T) & 1
\end{bmatrix}\begin{bmatrix}
z(i) \\ z'(i)
\end{bmatrix}\\
y_{i,+}&=: A(\sigma_i)y_i+B(\sigma_i)G_i\, .
\end{align*}
Now since $z,z'$ are synced for the two chains $y=(x,v), y'=(x',v')$ at each step, $y_+-y'_+ = A(y-y')$. Using $v_0=v_0' \sim \mathcal{N}(0,I)$ but $x_0\sim \rho_0 \neq x_0'\sim \pi_x$ after initialization
\begin{align*}
  W_2^2(\rho_K,\pi_x)\leq \mathbb{E}[\|x_K-x_K'\|^2] &\leq \mathbb{E}[\|y_K-y_K'\|^2]  \\
  &= \mathbb{E}[\|A^K(y_0-y_0')\|^2]\\
  &\leq \|A^{\top}A\|_{op}^K\cdot  \mathbb{E}[\|y_0-y_0'\|_2^2] = \rho(A^{\top}A)^K\cdot\mathbb{E}[\|x_0-x_0'\|_2^2]\\
  &= \rho(A^{\top}A)^K\cdot W_2^2(\rho_0,\pi_x)\, , 
\end{align*}
where the leftmost inequality simply used the definition of the Wasserstein distance and we took an infimum over the initial coupling of the positions. Taking the square root on both sides finish the argument. Since the entire transition matrix $A$ can be permuted to a block-diagonal matrix of size $2\times 2$ as shown above, we only need to look at $\sup_{\sigma_i\in(\mu,L)}\|A(\sigma_i)^{\top}A(\sigma_i)\|_{op}=\sup_{\sigma_i\in(\mu,L)}\rho(A(\sigma_i)^{\top}A(\sigma_i))$ the spectral radius of each block. 
\end{proof}
This investigation is close in spirit to \cite{monmarche2022hmc} but non-asymptotic in nature. They consider a sequence of problems with particular scaling $L_n h_n\rightarrow T$, $\eta_n\rightarrow \eta$ where $h_n\rightarrow 0$ is the stepsize in the numerical integrator, $L_n$ the number of integration steps within a refreshment interval and studied optimal choice of parameters $(\eta,T)$ in the limit as $n\rightarrow \infty$. 

\begin{lemma} \label{lem:spectrum}
The eigenvalues of $A(\sigma_i)^{\top}A(\sigma_i)$ are $\frac{1}{2}(b\pm \sqrt{b^2-4\eta^4})$, for
\[b:=\sin^2(\sqrt{\sigma_i}T)\eta^2(\sigma_i+\frac{1}{\sigma_i})+\cos^2(\sqrt{\sigma_i} T)(1+\eta^4)\geq 0\, .\]
Since $0\leq b^2-4\eta^4 \leq b^2$, the spectral radius $\rho(A(\sigma_i)^{\top}A(\sigma_i))$ is $\frac{1}{2}(b+ \sqrt{b^2-4\eta^4})$. 

\end{lemma}
\begin{proof}
Looking at the characteristic polynomial of 
\begin{align*}
&\det(A(\sigma_i)^{\top}A(\sigma_i)-r\cdot I_{2\times 2} )\\
&=\det\begin{bmatrix}
\cos^2(\sqrt{\sigma_i}T)+\eta^2\sigma_i\sin^2(\sqrt{\sigma_i}T)-r & (*)\\
\frac{\eta}{\sqrt{\sigma_i}}\sin(\sqrt{\sigma_i}T)\cos(\sqrt{\sigma_i}T)-\eta^3\sqrt{\sigma_i}\cos(\sqrt{\sigma_i}T)\sin(\sqrt{\sigma_i}T) & \frac{\eta^2}{\sigma_i}\sin^2(\sqrt{\sigma_i}T)+\eta^4\cos^2(\sqrt{\sigma_i}T)-r
\end{bmatrix} \\
&= 0
\end{align*}
for 
\[(*)=\frac{\eta}{\sqrt{\sigma_i}}\cos(\sqrt{\sigma_i}T)\sin(\sqrt{\sigma_i}T)-\eta^3\sqrt{\sigma_i}
\sin(\sqrt{\sigma_i}T)\cos(\sqrt{\sigma_i}T)\, ,\]
which gives
\[r^2-r\left(\sin^2(\sqrt{\sigma_i}T)[\eta^2\sigma_i+\eta^2/\sigma_i]+\cos^2(\sqrt{\sigma_i}T)[1+\eta^4]\right)+[\eta^2\cos^2(\sqrt{\sigma_i}T)+\eta^2\sin^2(\sqrt{\sigma_i}T)]^2=0\, .\]
Therefore the larger of the two roots is as claimed, using the quadratic formula, together with $\sigma+1/\sigma \geq 2$ and $1+\eta^4-2\eta^2=(1-\eta^2)^2\geq 0$.
\end{proof}

With this result, we aim to work out the optimal $\eta$ and $T$ that gives the largest contraction ratio $\rho$. But before proceeding, we entertain the possibility that had we only been able to pick $T$ and $\eta=0$, in this case $\rho=|\cos(\sqrt{\sigma_i}T)|$, so choosing $T=\frac{\pi}{\sqrt{L}+\sqrt{\mu}}$ solves 
\begin{equation}
\label{eqn:optimal_T}    
\min_T\, \max\{|\pi/2-\sqrt{\mu}T|,|\sqrt{L}T-\pi/2|\}
\end{equation}
which gives the best contraction. Now since $\frac{\pi\sqrt{\sigma_i}}{\sqrt{\mu}+\sqrt{L}}\in[0,\pi]$ for all $\sigma_i\in[\mu,L]$, using $\cos(x)\leq 1-\frac{1}{8}x^2$ for $x\in[0,\pi]$,
\[|\cos(\sqrt{\sigma_i}T)|\leq 1-\frac{1}{8}\sigma_iT^2\leq 1-\Theta\left(\left(\frac{\sqrt{\mu}}{\sqrt{L}+\sqrt{\mu}}\right)^2\right)=1-\Theta\left(\frac{\mu}{L}\right)=1-\Theta(1/\kappa)\,.\]
for all $i\in[d]$ and $k\in[K]$. In the above the two extreme endpoints $\mu$ and $L$ are symmetric, thanks to \eqref{eqn:optimal_T} and the symmetry property of $\cos(\cdot)$ around $\pi/2$. This recovers the result of \cite{chen_vempala2022optimal} in the Gaussian case for $\eta=0$ and $T = \pi/2\sqrt{L}$ with rate of $\mathcal{O}(\kappa\log(1/\epsilon))$ and suggests the choice of $T$ there is essentially optimal.  
\begin{remark} 
Dependence on $d$ only enters the picture when there's discretization error $\delta$; otherwise the rate is dimension-free. We also remark that a somewhat fairer metric would be $min\; K$ such that $ W_2(\rho_K,\pi_x)\leq \epsilon\sqrt{d/\mu}$ since this is the effective diameter of the problem and $W_2$ is an expansive quantity, but since various previous works have not adopted this convention and for easier comparison we will not use this alternative here.
\end{remark}
\begin{proposition}[Ideal HMC with partial refreshment]
\label{prop:partial_refresh}
For $f(x)=\frac{1}{2}x^{\top}\Sigma x$, Algorithm \ref{alg:momentum} ran with $\eta = \frac{1-\sin(\frac{\pi}{1+\sqrt{\kappa}})}{\cos(\frac{\pi}{1+\sqrt{\kappa}})}$, $T =\pi/(\sqrt{L}+\sqrt{\mu})$, after $K = \mathcal{O}(\sqrt{\kappa}\log(1/\epsilon))$ steps satisfies $W_2(\rho_K,\pi_x)\leq\epsilon$.
\end{proposition} 
\begin{proof}
Following Lemma \ref{lem:spectrum}, we rewrite the spectral radius and optimize $\eta$ given any fixed $T,\sigma_i$ (let $\sigma_i+1/\sigma_i=: 2/c(\sigma_i)$ for $c(\sigma_i)\in (0,1]$)
\[\min_{\eta\in(0,1)}\;\eta^2+\frac{D}{2}+ \frac{1}{2}\sqrt{(2\eta^2+D)^2-4\eta^4} = \min_{\eta\in(0,1)}\;\eta^2+\frac{D}{2}+\frac{1}{2}\sqrt{D(D+4\eta^2)}\]
where 
\[D:=\sin^2(\sqrt{\sigma_i}T)2\eta^2(\frac{1}{c(\sigma_i)}-1)+\cos^2(\sqrt{\sigma_i}T)(1-\eta^2)^2\,.\]
Note the first term in $D$ is increasing in $\eta$ and the second decreasing, we pick $\eta(\sigma_i,T)$ such that
\begin{equation}
\label{eqn:optimal-eta}    
\frac{2\eta^2}{(1-\eta^2)^2}= \frac{\cos^2(\sqrt{\sigma_i}T)}{2\sin^2({\sqrt{\sigma_i}T})}
\end{equation}
to minimize the sum, in which case $D=\sin^2(\sqrt{\sigma_i}T)2\eta^2(\frac{1}{c(\sigma_i)}+1)$. In broad strokes, this makes the $\eta^2+\sin^2(\cdot)\eta^2$ part appear so that we don't rely solely on the $\cos^2(\cdot)$ part (which recover the classical HMC as we saw).
Therefore the contraction ratio is dictated by
\[\min_T \max_{\sigma_i\in[\mu,L]}\; \eta^2+\sin^2(\sqrt{\sigma_i}T)\eta^2\left(\frac{1}{c(\sigma_i)}+1\right)\, .\] 
Imposing $\sin^2(\sqrt{\sigma_i}T)(\frac{1}{c(\sigma_i)}+1)=o(1)$, it implies $\eta^2\approx D$, which yields $\Theta(\eta^2)$, therefore contraction rate $\rho\asymp|\eta|$. To pick the best $T$, since from \eqref{eqn:optimal-eta},
\[|\eta|=\left|\frac{1-\sin(\sqrt{\sigma_i}T)}{\cos(\sqrt{\sigma_i}T)}\right|\]
is monotonically decreasing on $[0,\pi/2]$ and monotonically increasing on $[\pi/2,\pi]$, the same justification as before 
\[\min_T \; |\pi/2-\sqrt{\mu}T|\vee |\sqrt{L}T-\pi/2|\] 
draws the conclusion $T=\frac{\pi}{\sqrt{L}+\sqrt{\mu}}$ and $|\eta|\in[0,1)$. 

Using $\sin(x)\leq x$ for $x\in(0,\pi/2)$, we check $\left|\sin(\sqrt{\sigma_i}T)\sqrt{c(\sigma_i)^{-1}+1}\right|$ is orderwise upper bounded by (it suffices to check one of the endpoints $\mu$ and $L$ because of the symmetry, thanks to the choice of $T$)
\[\frac{1}{1+\sqrt{\kappa}}\times\{\sqrt{\mu+1/\mu} \vee \sqrt{L+1/L}\}\asymp \frac{\sqrt{\kappa}}{1+\sqrt{\kappa}}\rightarrow o(1) \quad \text{provided}\quad \kappa\rightarrow\infty\]
also that the corresponding rate approaches
\[\rho\asymp \frac{1-\sin(\frac{\pi}{1+\sqrt{\kappa}})}{\cos(\frac{\pi}{1+\sqrt{\kappa}})}\asymp 1-\sin\left(\frac{\pi}{1+\sqrt{\kappa}}\right)\asymp 1-\Theta\left(\frac{1}{\sqrt{\kappa}}\right)\]
for $\kappa$ large. This, together with Lemma \ref{lem:harmonic_oscillator}, render the $\mathcal{O}(\sqrt{\kappa}\log(1/\epsilon))$ iteration complexity. 
\end{proof}
We note that the choice of the integration length $T =\pi/(\sqrt{L}+\sqrt{\mu})$ here is longer by at most a factor of 2 compared to \cite{chen_vempala2022optimal}, and very close to $\mathcal{O}(1/\sqrt{L})$ for $\kappa$ large, which is shorter than Chebyshev integration length that would take some intervals to be as long as $\pi/2\sqrt{\mu}$. Hence the benefit seems to come mostly from the persistence parameter $\eta\rightarrow 1$ as $\kappa \rightarrow \infty$. The algorithm, however, only requires knowledge of $\mu$ and $L$.

\subsection{Randomized Integration Time}
As users of HMC may tell, integration time is notoriously hard to tune in practice, inspiring various proposals on randomizing this choice to hopefully alleviate the sensitivity to resonance effects. For Poisson jump process with intensity $1/\lambda$, the random times between successive jumps are independent and exponentially distributed with mean $\lambda$, which means $T_k \sim\text{Exp}(1/\lambda)$. RHMC is known to be more well-behaved in that it is geometrically ergodic as soon as $\lambda>0$ \cite{randomized_hmc} for e.g., strongly-convex \& smooth potentials, in contrast to constant time HMC, which requires $\eta$ small enough to ensure adequate randomness for non-Gaussian potentials. Curious connection to piece-wise deterministic Markov chain monte carlo (e.g., bouncy particle, zig-zag sampler) is subject of exploration in \cite{PDMP}. 
\begin{algorithm}[H]
\caption{Ideal Randomized HMC}\label{alg:rmhc}
\begin{algorithmic}
\Require Initial $(x_0,v_0)\in\mathbb{R}^d\times\mathbb{R}^d \sim \rho_0\otimes\mathcal{N}(0,I_d)$ independent 
\Require Poisson clock rate $1/\lambda$, friction $\eta\in[0,1)$, num of iterations $K>1$
\For{$k=1,\cdots,K$}
\State Draw independent $z\sim\mathcal{N}(0,I_d)$
\State Draw $T_k\sim\text{Exp}(1/\lambda)$ 
\State $(x_k,\tilde{v}_k) = \text{HMC}_{T_k}(x_{k-1},v_{k-1})$
\State $v_k=\eta \tilde{v}_k+\sqrt{1-\eta^2} \cdot z$
\EndFor
\State\Return $x_1,\cdots, x_K$
\end{algorithmic}
\end{algorithm} 
\begin{proposition}[Ideal RHMC]  
\label{prop:rhmc}
We have after $K=\Theta(\log(1/\epsilon))$ jumps with $\eta=0$ and $1/\lambda=2\sqrt{\mu}$ for Algorithm \ref{alg:rmhc} on $f(x)=\frac{1}{2}x^{\top}\Sigma x$, $W_2(\rho_K,\pi_x) \leq \epsilon$. In expectation, the total time elapsed is $\mathcal{O}(\frac{1}{\sqrt{\mu}}\log(1/\epsilon))$.
\end{proposition}
\begin{proof}
Since each $T_k$ is independent and identically distributed during the execution of the algorithm (independent of the state of the chain as well), and complete refreshment allows cancelling the stochasticity induced by $v, v'$ at each step, using Lemma \ref{lem:harmonic_oscillator}:
\begin{align*}
    W_2^2(\rho_K,\pi_x)\leq\mathbb{E}[\|x_K-x_K'\|^2] &\leq \mathbb{E}\left[\sum_{j=1}^d \left(\prod_{k=1}^K\cos(\sqrt{\sigma_j}T_k)\times(x_0(j)-x_0'(j))\right)^2\right]\\
    &=\sum_{j=1}^d \left(\prod_{k=1}^K\mathbb{E}\left[\cos^2(\sqrt{\sigma_j}T_k)\right]\right)\times \mathbb{E}\left[(x_0(j)-x_0'(j))^2\right]\\
    &\leq \left(\max_j\prod_{k=1}^K\mathbb{E}[\cos^2(\sqrt{\sigma_j}T_k)]\right)\times \mathbb{E}\left[\sum_{j=1}^d (x_0(j)-x_0'(j))^2\right]\\
    &\leq \left(\max_j\,\mathbb{E}[\cos^2(\sqrt{\sigma_j}T)]\right)^K\times W_2^2(\rho_0,\pi_x)\, .
\end{align*}
Taking square root on both sides conclude the argument therefore it remains to bound $\rho :=\max_j \mathbb{E}[\cos^2(\sqrt{\sigma_j}T)]$ for $T\sim\text{Exp}(1/\lambda)$. For this, we have for $\sigma_j\in[\mu,L]$
\begin{align*}
   &\frac{1}{\lambda}\int_0^\infty \exp(-\frac{1}{\lambda}T)\cos^2(\sqrt{\sigma_j}T) \, dT\\
   &=\frac{1}{\lambda}\int_0^\infty \exp(-\frac{1}{\lambda}T) \left(\frac{\exp(i\sqrt{\sigma_j}T)+\exp(-i\sqrt{\sigma_j}T)}{2}\right)^2 \, dT\\
   &=\frac{1}{4\lambda}\int_0^\infty \exp(-\frac{1}{\lambda}T) \sum_{j=0}^2 {2 \choose j}\exp(i\sqrt{\sigma_j}T)^{2-j}\exp(-i\sqrt{\sigma_j}T)^j\, dT\\
   &=\frac{1}{4\lambda}\Bigg(\frac{\exp(-\frac{T}{\lambda})(\cos(2\sqrt{\sigma_j}T)+i\sin(2\sqrt{\sigma_j}T))}{-1/\lambda+2i\sqrt{\sigma_j}}+\frac{2\exp(-T/\lambda)}{-1/\lambda}\\
   &\quad \quad \quad \quad \quad \quad +\frac{\exp(-\frac{T}{\lambda})(\cos(-2\sqrt{\sigma_j}T)+i\sin(-2\sqrt{\sigma_j}T))}{-1/\lambda-2i\sqrt{\sigma_j}}\Bigg)\biggr\rvert_{T=0}^\infty\\
   &=\frac{\exp(-T/\lambda)\left(-\cos(2\sqrt{\sigma_j}T)\cdot\frac{2}{\lambda}+\sin(2\sqrt{\sigma_j}T)\cdot 4\sqrt{\sigma_j}\right)\biggr\rvert_{T=0}^\infty}{4\lambda (-1/\lambda+2i\sqrt{\sigma_j})(-1/\lambda-2i\sqrt{\sigma_j})}+\frac{1}{\lambda}\times \frac{\lambda}{2}\\
   &=\frac{1}{2}+\frac{1}{2+8\lambda^2\sigma_j}=1-\frac{2\lambda^2\sigma_j}{1+4\lambda^2\sigma_j} \leq \exp\left(-\frac{2\lambda^2\sigma_j}{1+4\lambda^2\sigma_j}\right)
\end{align*}
where we used Euler's identity for complex exponentials, binomial theorem and symmetric properties of the sinusoids. Consequently, the total time duration (in expectation) is
\[\min_\lambda \max_{\sigma_j\in[\mu,L]}\; \lambda\cdot\frac{2+8\lambda^2\sigma_j}{2\lambda^2\sigma_j}\log(1/\epsilon)=\min_\lambda \max_{\sigma_j\in[\mu,L]}\; \left(\frac{1}{\lambda\sigma_j}+4\lambda\right)\log(1/\epsilon) =\frac{4}{\sqrt{\mu}}\log(1/\epsilon)\]
if picking $\lambda=\frac{1}{2\sqrt{\mu}}$. 
\end{proof}
This choice of $\lambda$ allows in expectation longer integration time $\Theta(1/\sqrt{\mu})$ closer to Chebyshev integration time. For a meaningful comparison, note for constant time \emph{without} damping, we have from \cite{chen_vempala2022optimal} the total duration is 
\begin{equation}
 \label{eqn:complexity}   
\frac{1}{\sqrt{L}}\times \frac{L}{\mu}\log(1/\epsilon) = \frac{\sqrt{L}}{\mu}\log(1/\epsilon)\,. 
\end{equation}
While for constant time \emph{with} damping, we have from Section \ref{sec:partial} that \[\frac{\pi}{\sqrt{\mu}+\sqrt{L}}\times \sqrt{\frac{L}{\mu}}\log(1/\epsilon)\approx \frac{1}{\sqrt{\mu}}\log(1/\epsilon)\]
for $\kappa$ large, which is $\sqrt{\kappa}$ factor faster. Additionally, the randomized integration time above also gives the same time (in expectation) as the damping-based method. The argument above also suggests one has some freedom in the choice of the distribution for the duration, but we will not explore this option here.

\begin{remark}
For quadratic potentials, one can show that the commonly employed velocity Verlet integrator with stepsize $h$ follows an \emph{exact} trajectory for a modified Hamiltonian $\tilde{\mathcal{H}}(x,v) = \frac{1}{2}\sum_i \tilde{\sigma}_i x_i^2+\frac{1}{2}\|v\|_2^2$ where $\tilde{\sigma}_i=\sigma_i(1-h^2\sigma_i/4)$ \cite{leimkuhler2004simulating}. Therefore if we know the convergence rate for the ideal HMC to its equilibrium w.r.t $\tilde{\mathcal{H}}$ (note the error only comes from the Hamiltonian discretization, partial/full refreshment does not introduce errors), since the energy difference $\mathcal{H}(x,v)-\tilde{\mathcal{H}}(x,v)=\frac{h^2}{8}\sum_{i=1}^d \sigma_i^2 x_i^2$, the guarantee for the unadjusted HMC follows immediately in this case by invoking Proposition \ref{prop:partial_refresh} and \ref{prop:rhmc} with $h=\mathcal{O}(\sqrt{\epsilon}/(\sqrt{L}d^{1/4}))$ and consequently the total number of gradient queries is roughly $1/(\sqrt{\mu}h) = \tilde{\mathcal{O}}(\sqrt{\kappa} d^{1/4}/\sqrt{\epsilon})$. 
\end{remark}

\subsection{Variable Integration Time}
We compare in what follows another variable integration time with complete refreshment proposed in \cite{wang2022accelerating} that has the same $\mathcal{O}(\sqrt{\kappa}\log(1/\epsilon))$ iteration complexity on quadratics as the damping method. For convenience, we reproduce the algorithm below. Some caveats of Chebyshev integration time worth bringing up include: (1) for practical purpose we don't usually know $K$ in advance; (2) the guarantee is only on the last iterate; (3) numerical stability is another concern when $\mu$ is small. These might be addressed by intermittent restarts but the accelerated rate is no longer assured.
\begin{algorithm}[H]
\caption{Ideal Chebyshev HMC \cite{wang2022accelerating}}\label{alg:chebyshev}
\begin{algorithmic}
\Require Initial $(x_0,v_0)\in\mathbb{R}^d\times\mathbb{R}^d \sim \rho_0\otimes\mathcal{N}(0,I_d)$ independent 
\Require Num of iterations $K>1$  
\State For a random shuffling of the $K$ indices $\sigma(k)\colon [K]\mapsto [K]$, set
\[T_{\sigma(k)}=\frac{\pi}{2\sqrt{L+\mu-(L-\mu)\cos(\frac{(\sigma(k)-1/2)\pi}{K})}}\]
\For{$k=1,\cdots,K$}
\State $(x_k,v_k) = \text{HMC}_{T_{\sigma(k)}}(x_{k-1},v_{k-1})$
\State Draw $z\sim\mathcal{N}(0,I_d)$, set $v_k\gets z$
\EndFor
\State\Return Last iterate $x_K$
\end{algorithmic}
\end{algorithm} 

\begin{corollary}[Guarantee for Chebyshev HMC]
Algorithm \ref{alg:chebyshev} outputs $x_K\sim\rho_K$ for $f(x) = \frac{1}{2} x^{\top}\Sigma x$ such that $W_2(\rho_K,\pi_x)\leq \epsilon$ in total time $\mathcal{O}(\frac{1}{\sqrt{\mu}}\log(1/\epsilon))$.
\end{corollary}
\begin{proof}
Recall the expression $T_k$ for Chebyshev is for $k\in[K]$, 
\[T_k=\frac{\pi}{2\sqrt{L+\mu-(L-\mu)\cos(\frac{(k-1/2)\pi}{K})}}=:\frac{\pi}{2}\frac{1}{\sqrt{2r_k}}\, .\]
Order does not play a role here (i.e., exchangeable) since the contraction rate only depends on the product \[\max_{j\in[d]}\,\left|\prod_{k=1}^K\cos\left(\frac{\pi\sqrt{\sigma_j}}{2\sqrt{r_k}}\right)\right|\,, \]
as made clear by Lemma \ref{lem:harmonic_oscillator}.
Putting $K=\sqrt{\kappa}\log(1/\epsilon)$ \cite{wang2022accelerating}, w.l.o.g assume $K$ is even, the roots $\{r_k\}_{k=1}^K$ form pairs, being equi-spaced points on the semicircle between $\mu$ and $L$. Taking derivatives of the function 
\[p(j):=\frac{1}{\sqrt{L+\mu-(L-\mu)\cos(\frac{(j-1/2)\pi}{K})}}+\frac{1}{\sqrt{L+\mu-(L-\mu)\cos(\frac{(k-j+1/2)\pi}{K})}}\]
for $j=1,\cdots,K/2$, we have for $b:=\sin(\frac{(j-1/2)\pi}{K})>0$ and $a:=\cos(\frac{(j-1/2)\pi}{K})\in(0,1)$,
\[p'(j)=\frac{b\pi}{2K}\left([L+\mu-(L-\mu)a]^{-3/2}-[(L+\mu+(L-\mu)a)]^{-3/2}\right) > 0\, ,\]
which means that $p(j)$ is an increasing function of $j$ therefore it suffices to look at the middle two indices closest to $\pi/2$. For this, setting $j=K/2$, since both $\frac{K-1}{K}$ and $\frac{K+1}{K}$ approach $1$ for $K$ (equivalently $\kappa$) large, we have the total duration of time $\sum_{k=1}^K T_k$ is upper bounded by
\[\mathcal{O}\left(\sqrt{\kappa}\log(1/\epsilon)\times\frac{1}{\sqrt{L+\mu}}\right)\approx \mathcal{O}\left(\frac{1}{\sqrt{\mu}}\log(1/\epsilon)\right)\]
for $\kappa$ large.
\end{proof}
One could view this as a ``de-randomized" version of RHMC where longer integration up to $\Theta(1/\sqrt{\mu})$ time is permitted, without being wasteful in making U-turns. While being quite neat, this comes at the cost of not being an any-time algorithm any more and the choice of the specific parameters are quite delicate/magical which we find it hard to believe as perhaps ``the right way to view things" (but this is more of a humble opinion, not a fact).

\section{Towards General Strongly-Convex, Smooth Potentials}
\label{sec:general}
For general strongly convex, smooth potentials where $\mu\cdot I\preceq \nabla^2 f(x)\preceq L\cdot I$, the result of \cite{dalalyan2020sampling} gave contraction rate of $\mathcal{O}(\mu/\sqrt{L + \mu})\approx \mathcal{O}(\mu/\sqrt{L})$ for the optimal choice of the friction $\gamma=\sqrt{\mu+L}$ for under-damped Langevin and \cite{chen_vempala2022optimal} gave total time $\mathcal{O}(\sqrt{L}/\mu \cdot \log(1/\epsilon))$ to equilibrium as we just saw in \eqref{eqn:complexity} for constant time HMC, both in continuous time without discretization for $W_2$. But the Gaussian example from Section \ref{sec:quadratic} points to the possibility of a $\sqrt{\kappa}$ gap between variable integration time / partial refreshment and constant integration time HMC.

On the other hand, over-damped Langevin only requires strong convexity for linear convergence with rate $\mathcal{O}
(\frac{1}{\mu}\log(1/\epsilon))$, although the benefit lies in the smoothness of the sample path when it comes to discretization for second-order dynamics. But for the sake of comparison, we see that HMC from Section \ref{sec:quadratic} could be advantageous too when $\mu$ is small (going from $1/\mu$ to $1/\sqrt{\mu}$ in continuous time).


Throughout this section, we will assume $\kappa\rightarrow \infty$, i.e., $L\gg \mu$. The theme is to bring to bear randomization of the integration time, together with partial momentum refreshment to ``add enough friction" to avoid rapid dissipation in the dynamics. We focus in this section on arguments that do not rely on analytical formulas for the dynamics as carried out in the previous section.  

\subsection{Jump Process}
\label{sec:jump}
A natural generalization would be the following random jumping time dynamics studied in \cite{malt2022}. Although the algorithm itself does not yield our desired rate, we sketch the argument provided there since it will be useful for our later discussion. To begin, we observe Algorithm \ref{alg:rmhc} is an implementation of the following jump process driven by the SDE: 
\begin{equation}
\label{eqn:jump_process}    
dx_t = v_t\, dt\quad\quad dv_t = -\nabla f(x_t)\, dt + \left(\sqrt{1-\eta^2}z_{N_{t-}}+(\eta-1)v_{t-}\right)\, dN_t\, .
\end{equation}
where $(N_t)_{t>0}$ is a time-homogeneous Poisson process with rate $\lambda^{-1}$ and $v_{t-}$ is the velocity immediately before the jump. The infinitesimal generator of the process acting on test function $g(x,v)\in \mathcal{C}^1(\mathbb{R}^{2d})$ is:
\begin{align*}
&\left(v^{\top}\nabla_x g(x,v)-\nabla f(x)^{\top}\nabla_v g(x,v)\right)+\lambda^{-1} \cdot \left(\mathbb{E}[g(x,\eta v+\sqrt{1-\eta^2}z)]-g(x,v)\right)\\
&=:\mathcal{L}_\text{Hamiltonian/Liouville}\, g(x,v)+ \lambda^{-1}\cdot \mathcal{L}_{\text{poisson-process},\eta}\, g(x,v)\, ,
\end{align*}
which has an energy-preserving Hamiltonian component and an additional part that is related to the Fluctuation-Dissipation theorem from statistical mechanics. Now if applying synchronous coupling (both for the clock and the refreshment $z$) to two processes $y_t = (x_t,v_t), y_t' = (x_t',v_t')$ following dynamics \eqref{eqn:jump_process}, we have for $\tilde{x}_t:=x_t-x_t'$, $\tilde{v}_t:=v_t-v_t'$ the two coupled process evolve as
\[d\tilde{x}_t = \tilde{v}_t\, dt\quad\quad d\tilde{v}_t = -H_t\tilde{x}_t \, dt + (\eta-1)\tilde{v}_t\, dN_t\]
where we used mean value theorem for 
\[\mu I\preceq H_t:=\int_0^1 \nabla^2f(sx_t+(1-s)x_t')\, ds \preceq L I.\]
At this point, if we track an extended Lyapunov-function-type quantity, which is common in works on second-order dynamics \cite{dalalyan2020sampling}: $\|y_t-y_t'\|_A^2 = a\|\tilde{x}_t\|^2+2b\tilde{x}_t^{\top}\tilde{v}_t + c\|\tilde{v}_t\|^2$ for
\[A:=\begin{bmatrix}
aI_d & bI_d\\
bI_d & cI_d
\end{bmatrix} \succ 0\, .\]
Applying It\^{o}'s product rule from stochastic calculus, 
\begin{align*}
    \mathcal{L}\, \|y_t-y_t'\|_A^2 &= 2a\tilde{x}_t^{\top}\tilde{v}_t + 2b\tilde{x}_t^{\top}\left(-H_t\tilde{x}_t+(\eta-1)\lambda^{-1} \tilde{v}_t\right)+2b\|\tilde{v}_t\|^2\\
    &+2c\left(-\tilde{x}_t^{\top}H_t +(\eta-1)\lambda^{-1}\tilde{v}_t^{\top}\right)\tilde{v}_t+c(\eta-1)^2\lambda^{-1}\|\tilde{v}_t\|^2 =: -(y_t-y_t')^{\top}S_t(y_t-y_t')
\end{align*}
for 
\[S_t:=\begin{bmatrix}
2bH_t & (b(1-\eta)\lambda^{-1}-a)I_d+cH_t\\
(b(1-\eta)\lambda^{-1}-a)I_d+cH_t & (c(1-\eta^2)\lambda^{-1}-2b)I_d
\end{bmatrix}\, .\]
So if we can have $S_t \succeq 2rA$, then we deduce exponential convergence from Gr\"onwall's lemma:
\[\mathcal{L}\, \|y_t-y_t'\|_A^2  \leq -2r\|y_t-y_t'\|_A^2\Rightarrow \|y_T-y_T'\|_A\leq e^{-rT} \|y_0-y_0'\|_A\,.\]
To satisfy the requirement $S_t\succeq 2rA\succ 0\, \forall t$, it entails $\forall \sigma\in[\mu,L]$ eigenvalues of $H_t$, 
\begin{align}
\label{eqn:requirement1}
    a,c > 0, ac-b^2&>0\\
    -2ra+2b\sigma &\geq 0\\
     -2rc+c(1-\eta^2)\lambda^{-1}-2b & \geq 0\\
     \label{eqn:requirement4}
  [-2rb+b(1-\eta)\lambda^{-1}-a+c\sigma]^2&\leq [-2ra+2b\sigma][-2rc+c(1-\eta^2)\lambda^{-1}-2b]\, .
\end{align}
Since after permutation, a $2\times 2$ matrix is positive-definite iff the two diagonals are and the determinant is. 

From here, the authors of \cite{malt2022} showed that for all $\eta\in[0,1)$ the optimal choice of $\lambda^{-1}=\frac{2\sqrt{L+\mu}}{1-\eta^2}$ gives contraction rate $r=\frac{(1+\eta)\mu}{2\sqrt{L+\mu}}$. Therefore the performance improves as $\eta\rightarrow 1$ and refresh rate $\lambda^{-1}\rightarrow \infty$. This limiting behavior in fact coincides with under-damped Langevin and the conclusion there is that under-damped Langevin match those of \eqref{eqn:jump_process} with optimal choice of parameters, since more frequent and partial refreshments are always helpful for \eqref{eqn:jump_process}. Stronger equivalence in terms of generators is also established in \cite{malt2022}.

Close inspection suggests their conclusion is not surprising, as in this case $\eta$ and $\lambda^{-1}$ are tied such that dissipation $\propto \lambda^{-1}(1-\eta^2)=2\sqrt{L+\mu}\asymp \sqrt{L}\gg \sqrt{\mu}$, which means there's only one degree of freedom in the dynamics (i.e., one could re-parametrize w.r.t $\lambda^{-1}(1-\eta)^2$ and essentially get similar result). This dissipation of $\approx \sqrt{L}$ is also what one gets with constant time integration (since in that case $\eta=0$, and one could think of $\lambda^{-1}\approx \sqrt{L}$). Note that picking $A = \begin{bmatrix}a\cdot \Sigma & b\cdot I \\ b\cdot I & c\cdot I\end{bmatrix}$ the argument above will recover the $\mathcal{O}(1/\sqrt{\mu})$ rate for quadratics with $\lambda^{-1}=\mathcal{O}(\sqrt{\mu})$ and $\eta=0$, but of course this is not generalizable beyond the constant Hessian case. 

\subsection{Randomized Coordinate Partial Refreshment}
To introduce more degrees of freedom in the choice of the parameters, another avenue to incorporate partial refreshment is coordinate-wise update, where each coordinate is assigned a different clock rate. It is clear that the unique invariant measure is still $\pi(x,v)\propto e^{-f(x)-\frac{1}{2}\|v\|^2}$ for quadratics. The reason for this delegation is if we look at the requirements \eqref{eqn:requirement1}-\eqref{eqn:requirement4}, given the opportunity to choose different $\lambda_i^{-1}$ for each $\sigma$, and without loss of generality, set $\eta=0$, we would be able to afford a more aggressive choice of $r$. This is made precise below. 
\begin{algorithm}[H]
\caption{Ideal Randomized Coordinate HMC}\label{alg:hmc_coordinate}
\begin{algorithmic}
\Require Initial $(x_0,v_0)\in\mathbb{R}^d\times\mathbb{R}^d \sim \rho_0\otimes\mathcal{N}(0,I_d)$ independent 
\Require Poisson clock rate $1/\{\lambda_i\}_{i=1}^d$, friction $\eta\in[0,1)$, time duration $T > 0$
\For{$s\in[0,T]$}
\State Continuously draw $\delta t_i\sim\text{Exp}(1/\lambda_i)$ for $i \in [d]$ independently
\If{Any clock for coordinate $i\in[d]$ rings}
\State Draw independent $z\sim\mathcal{N}(0,1)$ \State Update $v_s(i)\gets\eta v_s(i)+\sqrt{1-\eta^2} \cdot z$ while keeping other coordinates $j\neq i$ same
\Else \State Evolve HMC from current state $(x_s,v_s)$ forward according to \eqref{eqn:hmc}
\EndIf
\EndFor
\State\Return $x_T$
\end{algorithmic}
\end{algorithm} 

\begin{proposition}[Mixing of Randomized Coordinate HMC]
\label{prop:coordinate}
For potential $f(x)=\frac{1}{2}x^{\top}\Sigma x$ where $\Sigma=\text{diag}(\sigma_i)$ and $\mu\leq \{\sigma_i\}\leq L$, Algorithm \ref{alg:hmc_coordinate} with $\{\lambda_i^{-1}\} = \{\sqrt{\mu},\cdots,\frac{\sigma_i}{\sqrt{\mu}},\cdots,\frac{L}{\sqrt{\mu}}\geq\sqrt{L}\}$ and $\eta=0$ gives $W_2(\rho_T,\pi_x)\leq \epsilon$ in time $T=\mathcal{O}(\frac{1}{\sqrt{\mu}}\log(1/\epsilon))$.
\end{proposition} 
\begin{proof}
Keep in mind that $H_t=\nabla^2f(x) = \Sigma = \text{diag}(\sigma_i)$ for quadratics. Repeating the reasoning from Section \ref{sec:jump}, we find with independent clock for each coordinate with rate $\lambda_i^{-1}$ and $\eta=0$ (below $D_\lambda$ is diagonal matrix with $\lambda_i^{-1}$ on the diagonal), the $S_t \succeq 2rA$ condition translates to
\[\begin{bmatrix}
2bH_t & (bD_\lambda-a)I_d+cH_t\\
(bD_\lambda-a)I_d+cH_t & (cD_\lambda-2b)I_d
\end{bmatrix}-2r\begin{bmatrix}
aI_d & bI_d\\
bI_d & cI_d
\end{bmatrix} \succeq 0\, ,\]
as we are essentially replacing $\tilde{v}_t\, dN_t$ with $\sum_i \tilde{v}_t^{(i)}\, dN_t^{(i)}$ where $\tilde{v}_t^{(i)}$ has a single nonzero element $\tilde{v}_t(i)$ at location $i$ and $N_t^{(i)}$ has rate $\lambda_i^{-1}$. Equation \eqref{eqn:requirement1}-\eqref{eqn:requirement4} then becomes for each $\sigma\in[\mu,L]$:
\begin{align}
    a,c > 0, ac-b^2&>0\\
    -2ra+2b\sigma &\geq 0\\
     -2rc+c\lambda_i^{-1}-2b & \geq 0\\
  [-2rb+b\lambda_i^{-1}-a+c\sigma]^2&\leq [-2ra+2b\sigma][-2rc+c\lambda_i^{-1}-2b]\, .
\end{align}
It is not hard to see that with the choices 
\[b \asymp \sqrt{\mu},\, a\asymp \mu,\, c=o(1),\, \lambda_i^{-1} \asymp \frac{\sigma}{\sqrt{\mu}}\]
all conditions can be met with $r\asymp\sqrt{\mu}$. Now to finish the argument, we start from a particular coupling of $(x_0,v_0)\sim \rho_0\otimes \mathcal{N}(0,I)$ and $(x_0',v_0')\sim \pi_x \otimes \mathcal{N}(0,I)$ with synced Gaussian refreshment and Poisson clock. In particular $v_0=v_0'$ but $x_0\neq x_0'$. We just showed for a coupled $(y_t,y_t')$ following the dynamics, for all $T>0$ 
\[\mathcal{L}\, \|y_t-y_t'\|_A^2  \leq -2r\|y_t-y_t'\|_A^2\quad \Rightarrow \quad \mathbb{E}[\|y_T-y_T'\|_A^2]\leq e^{-2rT} 
\mathbb{E}[\|y_0-y_0'\|_A^2]  \,.\]
Now to compare the twisted-$A$-norm to the Euclidean norm $\|\cdot\|_2$, 
\[c(a\|x\|^2+2bx^{\top}v+c\|v\|^2)=(ac-b^2)\|x\|^2+\|bx+cv\|^2\geq (ac-b^2)\|x\|^2\,.\]
Therefore 
\[\mathbb{E}[\|x_T-x_T'\|_2^2]\leq \frac{c}{ac-b^2}\mathbb{E}[\|y_T-y_T'\|_A^2]\leq \frac{ce^{-2rT} }{ac-b^2}
\mathbb{E}[\|y_0-y_0'\|_A^2]=\frac{ac\cdot e^{-2rT} }{ac-b^2}
\mathbb{E}[\|x_0-x_0'\|_2^2]\]
where we used $v_0=v_0'$ for the last step.
Taking infimum over all initial couplings $(x_0,x_0')\sim\rho_0\otimes \pi_x$ and using the definition of Wasserstein-2 distance,
\[W_2(\rho_T,\pi_x)\leq e^{-rT}\sqrt{\frac{ac }{ac-b^2}}W_2(\rho_0,\pi_x)\]
since $x_T'\sim\pi_x$ for all $T\geq 0$. Plugging in the choices made before, we have after time $T=\mathcal{O}(\frac{1}{\sqrt{\mu}}\log(1/\epsilon))$, $W_2(\rho_T,\pi_x)\leq \epsilon$, as announced.
\end{proof}


We do not deem it as a satisfying algorithm for implementation, as it requires knowledge of the full spectrum, but it indicates that more fine-grained control on the dissipation is helpful. From this perspective, what we described in the previous section are in fact the two extreme realizations, both maintaining the same $\sqrt{\mu}$ dissipation --
in the sense Algorithm \ref{alg:momentum} has \[\lambda^{-1}(1-\eta^2)\approx \sqrt{L}\times \left(1-\left(1-\frac{\sqrt{\mu}}{\sqrt{L}}\right)^2\right)\approx \sqrt{L}\times \frac{\sqrt{\mu}}{\sqrt{L}}\approx \sqrt{\mu}\]
and Algorithm \ref{alg:rmhc} has (Algorithm \ref{alg:chebyshev} and \ref{alg:hmc_coordinate} also fall on this end of the spectrum)
\[\lambda^{-1}(1-\eta^2)\approx \sqrt{\mu}\times 1\] so one saturates $\eta\rightarrow 1$, and the other $\eta\rightarrow 0$, which explains their preference over classical HMC (that has dissipation $\sqrt{L}\gg \sqrt{\mu}$). It is likely if one could come up with other methods to implement dissipation at this level, the resulting algorithm will enjoy better rate as well.

The statement of Proposition \ref{prop:coordinate} is not surprising in the separable case, but a naive scheme of picking a diagonal $D_\eta$ instead of $\eta$ to allow for different dissipation wouldn't work for general potentials because of the changing $H_t$ -- nevertheless it suggests a heuristic scheme of estimating local curvature at the current position and setting $D_\eta$ accordingly in practice may offer possible speed-up. Simulating state-dependent jump rate can be more challenging but techniques such that Poisson thinning could be employed if one has knowledge of an upper bound on the rate.


\subsection{Damping}
An orthogonal direction to move beyond quadratics could be to have deterministic integration time, but with friction added, as in Algorithm \ref{alg:momentum}. One cautionary word, though, is that such partially-refreshed dynamics, when discretized and Metroplized, in order to satisfy the (generalized) detailed balance condition, would lead to velocity flip when rejected therefore backtracking on the progress made, which is invisible for its fully-refreshed counterpart. We make the observation that for $R$ (resp. $H$) the refreshment (resp. Hamiltonian) operation, since 
\[(R^{1/2}HR^{1/2})^K = R^{1/2}(HR)^{K-1}HR^{1/2}\,,\]
the first $R$ is invisible and the last $R$ doesn't affect the position output, it suffices to study contraction property of $HR$, which also decouples the choice of $\eta$ and $\lambda$ (or $T$ in this context). Again consider synchronous coupling (same $z$ and $z'$) and using mean value theorem, for a matrix $\mu I \preceq M_t\preceq LI$, let $O$ and $Q_t$ denote the symmetric and skew-symmetric matrices
\[O:=\begin{bmatrix}
I & 0 \\ 0 & \eta I
\end{bmatrix}\quad\quad Q_t:=\begin{bmatrix}
0 & I \\ -M_t & 0
\end{bmatrix}\, ,\]
we can write the deterministic dynamics for the coupled $(x,v)$, $(x',v')$ processes as 
\[\begin{bmatrix}
x_k-x_k'\\
v_k-v_k'
\end{bmatrix} =  OH_{T,k}\begin{bmatrix}
x_{k-1}-x_{k-1}'\\
v_{k-1}-v_{k-1}'
\end{bmatrix} \quad \quad \frac{d}{dt}\begin{bmatrix}
x_t-x_t'\\
v_t-v_t'
\end{bmatrix} =  Q_t\begin{bmatrix}
x_t-x_t'\\
v_t-v_t'
\end{bmatrix}\, .\]
Here $H_{T,k}$ denotes the Hamiltonian flow map (i.e., the right equation in the above display integrated for time $T$ starting at iterates from $k-1$). Note we cannot directly apply the classical (fully refreshed) HMC argument since we can not assume the velocity are synced for the 2 chains after each resampling. But the result does carry over to the first iteration since we can assume $x_0\neq x_0'$ with $v_0=v_0'\sim\mathcal{N}(0,I)$, and reuse the same Lyapunov function idea from Section \ref{sec:jump} to track $\|y_t-y_t'\|_A^2$ for $A\succ 0$. As a preliminary step, leveraging the differential inequalities in \cite{chen_vempala2022optimal} (c.f. proof of Lemma 6), we can verify for $T\leq \frac{1}{2\sqrt{L}}$, 
\begin{align}
\label{eqn:vempala}
\|x_T-x_T'\|^2 &\leq (1-\frac{\mu}{16L})\|x_0-x_0'\|^2\\
\|v_T-v_T'\|^2 &\leq \eta^2 \frac{L}{4}\|x_0-x_0'\|^2\\
\langle x_T-x_T',v_T-v_T'\rangle &\leq -\eta\frac{\mu}{2\sqrt{L}}\|x_0-x_0'\|^2\, .
\end{align}
Setting the damping $\eta=1-\frac{1}{\sqrt{\kappa}}\in[0,1)$, as what the quadratic case taught us, we are able to find a positive-definite matrix $A$ such that the relationship  
\begin{equation}
 \label{eqn:recurse}   
\|y_T-y_T'\|_A^2 \leq \left(1-\Theta\left(\frac{1}{\sqrt{\kappa}}\right)\right)\|y_0-y_0'\|_A^2\leq\exp\left(-\Theta\left(\frac{1}{\sqrt{\kappa}}\right)\right)\cdot a\|x_0-x_0'\|_2^2
\end{equation}
holds when $\kappa$ is large (i.e., $\sqrt{\frac{\mu}{L}}\gg \frac{\mu}{L}$). The result of \cite{chen_vempala2022optimal}, of course, only exploited \eqref{eqn:vempala}. If successive refreshment intervals also obey similar recursion \eqref{eqn:recurse} one would end up with the desired $\mathcal{O}(\frac{1}{\sqrt{\mu}}\log(1/\epsilon))$ rate -- we leave such pursuit as future work.




\section{Numerics}
We use symplectic leapfrog (i.e., position Verlet) integrator for simulating the Hamiltonian dynamics, which is known to be second-order accurate. Having the feature of (1) time-reversible (in physics sense, up to sign flip of the velocity); (2) volume-preserving in phase space, it allows simulating long trajectories without incurring too much error from the continuous dynamics. The update from $(x_0,v_0)$ to $(x_1,v_1)$ with stepsize $h$ is
\begin{equation}
 \label{eqn:leapfrog}   
x_{1/2} = x_0+\frac{h}{2}v_0 \quad\quad v_1 = v_0-h\nabla f(x_{1/2})\quad\quad x_1 = x_{1/2}+\frac{h}{2} v_1
\end{equation}
and the total number of gradient queries are $TK/h$. It is not hard to see that for $K=1$ and $\eta=0$, \eqref{eqn:leapfrog} reduces to a discretization of the over-damped Langevin SDE with stepsize $h^2/2$. Rewriting \eqref{eqn:leapfrog}, position Verlet takes the form of
\[x_{1}=x_0+\frac{h}{2}v_{0}+\frac{h}{2}v_{1}= x_{1/2}+\frac{h}{2}v_{1}\]  
\[v_{1}=v_0-\frac{h}{2}\nabla f(x_{1/2})-\frac{h}{2}\nabla f(x_{1/2})=v_0-h\nabla f(x_0+\frac{h}{2}v_0)\, ,\]  
which has the interpretation of performing a symmetrized explicit, followed by implicit update for the $x$ variable, and an implicit followed by explicit update for the $v$ variable; or in other words, the combination of trapezoidal and the implicit midpoint rule for the two variables. 

For the Gaussian case, it's known that the asymptotic bias in Wasserstein-2 distance \cite{monmarche2022hmc} scales with stepsize (and only stepsize $h$) as  \[\sqrt{d}\left(\frac{1}{\sqrt{L}}-\sqrt{\frac{1}{L}-\frac{h^2}{4}}\right)\quad \text{so we pick} \quad  h\approx \frac{\sqrt{\epsilon}}{(Ld)^{1/4}} \,,\]
for $\epsilon< \sqrt{d/L}$ and do not perform Metropolis-Hastings adjustment. We evaluate the effective sample size (ESS) as a measure of correlation between samples, which is computed as $K/(1+2\sum_k \gamma(k))$ where $\gamma(k)$ estimates the autocorrelation at lag $k$ and $K$ is the length of the chain (i.e., number of samples). To aggregate for the multivariate distribution, we record the min and mean ESS across the dimensions. The experiment is repeated 50 times for each method, for which the average statistics are reported below. Target accuracy is set to be $\epsilon=10^{-2}$.

Concentration of empirical covariance is calculated as $\|\hat{\Sigma}-\Sigma\|_F/\|\Sigma\|_F$ where $\hat{\Sigma}$ is the empirical average computed on the $50$ independent chains from their last samples. The chain length is set to be $K=2000$ for all algorithms and $T$ is prescribed by each algorithm. 
\begin{table}[H]
\label{tab:quadratic}
\centering
\begin{tabular}{c|| c| c| c} 
 \hline
 Algorithm & Min ESS & Mean ESS & Empirical Cov Error \\ [0.5ex] 
 \hline\hline
 Constant Time \cite{chen_vempala2022optimal} & 12.83 & 42.13 & 0.43 \\ 
 Chebyshev Time \cite{wang2022accelerating} (Alg \ref{alg:chebyshev}) & 35.78 & 124.99 & 0.41 \\
 HMC with damping (Alg \ref{alg:momentum}) & 41.57 & 133.03 & 0.53 \\
 RHMC (Alg \ref{alg:rmhc})  & 25.04 & 75.82 & 0.51 \\ 
 \hline
\end{tabular}
\caption{Result on quadratic potential ($\Sigma = \text{diag}(1,\cdots,d)$, $\mu=0, d=10$)}
\end{table}

In the presence of multi-scale data, one could benefit from multi-time-stepping methods from molecular dynamics in which different time steps are employed for a frequency splitting of the forces, where the potential $\nabla f(x) = \nabla f_1(x) + \nabla f_2(x)$ for which $f_1$ is fast varying and $f_2$ slow varying (but otherwise not necessarily separable, or decomposable into gradients, e.g., $\nabla f(x) = \sum_i  \langle \nabla f(x), e_i\rangle e_i$ is also viable). Written in Trotter splitting form it would read as $O(\bar{B}A\bar{B} )^K O$, where for some parameters $(\eta,h,\delta,K)$: 
\[O: v\leftarrow \eta v+\sqrt{1-\eta^2}z\]
\[A:x\leftarrow x+\delta \cdot v\]
\[\bar{B}: \text{perform}\; v\leftarrow v-\delta/2\cdot \nabla f_1(x) \;\text{if}\; t\neq 1, K\;\text{where}\; \delta K = h \]
\[\quad \quad \text{perform}\;v\leftarrow v-\delta/2\cdot \nabla f_1(x)-h/2\cdot \nabla f_2(x)\; \text{for} \; t=1,K\, .\]
The largest permissible stepsize, however, is still limited by the highest oscillatory component, but the periodic gradient evaluations for the lower frequency component at the beginning and end of each interval only can offer computational savings.



\section{Randomized Midpoint ODE Discretization}
In \cite{bou2022unadjusted}, the authors discovered that for target with Lipschitz gradient without assuming higher-order regularity, a randomized integrator is a better choice compared to the Verlet scheme, which has accuracy $\mathcal{O}(h^{3/2})$ vs. $\mathcal{O}(h)$. In this section, we investigate the accuracy of the proposed sMC integrator \cite{bou2022unadjusted} and some related variants for the quadratic potential (which has Lipschitz Hessian).



Since the potential is separable it suffices to track the 1D dynamics for the velocity Verlet integrator. For the potential $\frac{1}{2}\lambda^{2} x^2$, assume we start at stationary $x_0\sim\mathcal{N}(0,\lambda^{-2}), v_0\sim\mathcal{N}(0,1)$ independent of each other and we aim to quantify how much it deviates from the exact trajectory after running leapfrog with stepsize $h$. The update can be written as
\begin{equation}
\begin{bmatrix}
\label{eqn:leapfrog-1}
x_{1}\\ v_{1} \end{bmatrix}
= \begin{bmatrix}
1-\frac{h^2}{2}\lambda^2 & h\\ -h\lambda^2(1-\frac{h^2\lambda^2}{4}) & 1-\frac{h^2}{2}\lambda^2
\end{bmatrix} \begin{bmatrix}
x_{0}\\ v_{0} \end{bmatrix}
\end{equation}
compared to the exact flow
\[\begin{bmatrix}
\bar{x}_{1}\\ \bar{v}_{1} \end{bmatrix}
= \begin{bmatrix}
\cos(\lambda t) & \frac{1}{\lambda}\sin(\lambda t)\\ -\lambda\sin(\lambda t) & \cos(\lambda t)
\end{bmatrix} \begin{bmatrix}
x_{0}\\ v_{0} \end{bmatrix} \sim \begin{bmatrix}\mathcal{N}(0,\cos^2(\lambda t)\lambda^{-2}+\frac{1}{\lambda^2}\sin^2(\lambda t))\\ \mathcal{N}(0,\lambda^2\sin^2(\lambda t)\lambda^{-2} + \cos^2(\lambda t))
\end{bmatrix}
\]
that stays stationary forever. We carry out one update for this analysis, since one can check that the additional terms are lower order if the propagator is applied repeatedly. It is clear that
\[x_1\sim\mathcal{N}(0,(1-\frac{h^2}{2}\lambda^2)^2\lambda^{-2}+h^2)\]
which means the Wasserstein bias between the two Gaussians is
\begin{align*}
W_2(x_1,\bar{x}_1)&=\mathbb{E}\left[\left|\sqrt{(1-\frac{h^2}{2}\lambda^2)^2\lambda^{-2}+h^2}z-\lambda^{-1}z\right|^2\right]^{1/2}\\
&= \left|\sqrt{(1-\frac{h^2}{2}\lambda^2)^2\lambda^{-2}+h^2}-\lambda^{-1}\right|\\
&= \left|\sqrt{\frac{h^4}{4}\lambda^2+\frac{1}{\lambda^2}}-\frac{1}{\lambda}\right|=\mathcal{O}(h^2 \lambda )\, .
\end{align*}
This is easily extendable to $d$-dimensional Gaussian where we will have an extra $\sqrt{d}$ factor. On the other hand, for the stratified Monte Carlo (sMC) time integrator, the update is for some $\tau\sim\text{Unif}(0,h)$:
\begin{align}
\label{eqn:smc_1}
x_1 &= x_0+hv_0-\frac{h^2}{2}\nabla f(x_0+\tau v_0)\\
\label{eqn:smc_2}
v_1 &= v_0-h\nabla f(x_0+\tau v_0)
\end{align}
which when put into propagator form becomes 
\[\begin{bmatrix}
x_{1}\\ v_{1} \end{bmatrix}
= \begin{bmatrix}
1-\frac{h^2}{2}\lambda^2 & h-\frac{h^2}{2}\lambda^2 \tau\\ -h\lambda^2 & 1-h\lambda^2 \tau
\end{bmatrix} \begin{bmatrix}
x_{0}\\ v_{0} \end{bmatrix}
\]
therefore
\[x_1\sim\mathcal{N}(0,(1-\frac{h^2}{2}\lambda^2)^2\lambda^{-2}+(h-\frac{h^2}{2}\lambda^2\tau)^2)\]
so the expected variance of $x_1$ is
\begin{align*}
\frac{1}{h}&\int_0^h (1-\frac{h^2}{2}\lambda^2)^2\lambda^{-2}+(h-\frac{h^2}{2}\lambda^2\tau)^2\, d\tau\\
&= \frac{h^4}{4}\lambda^2+\frac{1}{\lambda^2}-\frac{h^4\lambda^2}{2}+\frac{h^6\lambda^4}{4}=\mathcal{O}(h^6) =: \text{var(leapfrog)}-\frac{h^4\lambda^2}{2}+\frac{h^6\lambda^4}{4}\, ,
\end{align*}
which in terms of Wasserstein bias is one order higher.
We interpret the sMC algorithm \eqref{eqn:smc_1}-\eqref{eqn:smc_2} as a randomized midpoint method for the integral formulation of the Hamiltonian solution \eqref{eqn:hmc}:
\begin{equation}
\label{eqn:x_exact}
x_h = x_0+hv_0-\int_0^h\int_0^s \nabla f(x_t)\, dt ds=x_0+hv_0-\int_0^h \nabla f(x_t)(h-t)\, dt
\end{equation}
\[v_h = v_0-\int_0^h \nabla f(x_t) dt\, ,\]
which performs an Euler update to approximate a midpoint solution $x_\tau, \tau\in(0,h)$ first and proceed to form an ``unbiased" estimator for the integrals. Driven by this, we are curious if (1) nested approximation; (2) symmetrized update can be brought to bear for further improvement in the Hessian Lipschitz case. More specifically, we consider given $(x_0,v_0)$
\begin{enumerate}
\item For $\tau \sim \text{Unif}(0,h)$:
\begin{align*}
x_{\tau,1} &= x_0+\tau v_0-\frac{\tau^2}{2} \lambda^2 x_0\\
x_{\tau,2} &= x_0+\tau v_0-\frac{\tau^2}{2}\lambda^2x_{\tau,1}\\
x_1 &= x_0+hv_0-h(h-\tau)\lambda^2x_{\tau,2}\\
v_1 &= v_0-h\lambda^2 x_{\tau,2}
\end{align*}
\item For $\tau \sim \text{Unif}(0,h)$ (Verlet will simply set $\tau=0$ deterministically):
\begin{align*}
v_{1/2} &= v_0 - \frac{h}{2} \cdot \lambda^2(x_0+\tau v_0)\\
x_{1} &= x_0+\frac{h}{2}v_{1/2}+\frac{h}{2}v_{1/2}\\
&=x_0+hv_0-\frac{h^2}{2}\lambda^2 (x_0+\tau v_0)=(1-\frac{h^2}{2}\lambda^2)x_0+(h-\frac{h^2\lambda^2}{2}\tau)v_0\\
v_{1} &= v_{1/2}-\frac{h}{2} \cdot\lambda^2 (x_{1}-\tau v_{1/2})\\
&= (1+\frac{h}{2}\lambda^2\tau) v_{1/2}-\frac{h}{2}\lambda^2 x_1
\end{align*}

\end{enumerate}
The motivation for (1) above is that one can view \eqref{eqn:x_exact} as solving a fixed point iteration in the sense of $x_t^*=\mathcal{T}(x_t^*)$ for the optimal curve $(x_t^*)_t$ and an operator $\mathcal{T}$, therefore by fixed point theorem composing the map $\mathcal{T}(\mathcal{T}(...\mathcal{T}(x_0)))$ more times should yield better approximation to the solution $x_t^*$. The second method above is motivated from symmetrizing the sMC integrator, since \eqref{eqn:smc_1}-\eqref{eqn:smc_2} is closer to ``randomized Euler" than ``randomized Leapfrog", so this version recruits an antithetic variate that is more in line with the trapezoidal rule. 

We have carried out detailed calculation for both, but neither confers a higher order than the sMC integrator \eqref{eqn:smc_1}-\eqref{eqn:smc_2} from this one-step analysis. This is, of course, not conclusive as to what the best integrator for HMC could be, but it is perhaps not entirely obvious if one could improve upon the proposal in \cite{bou2022unadjusted} further, under higher order smoothness, unlike what seems to be alluded to there.

In fact, another way to see the difference between sMC algorithm \eqref{eqn:smc_1}-\eqref{eqn:smc_2} and the leapfrog update \eqref{eqn:leapfrog-1} more generally is that \eqref{eqn:leapfrog-1} can be un-rolled as
\begin{align*}
v_n&=(x_{n+1}-x_{n-1})/2h \\
x_{n+1}-2x_n+x_{n-1} &= -h^2 \nabla f(x_n)
\end{align*}
whereas \eqref{eqn:smc_1}-\eqref{eqn:smc_2} admit the three-term recursion
\begin{align*}
x_{n+1}&=x_{n-1}+2h v_n+h^2/2 \nabla f(x_{[n-1,n]})-h^2/2 \nabla f(x_{[n,n+1]})\\
\Leftrightarrow (x_{n+1}-x_{n-1})/2h &=  1/2(v_n+h/2 \nabla f(x_{[n-1,n]}))+1/2(v_n-h/2 \nabla f(x_{[n,n+1]}))\\
x_{n+1}-2x_n+x_{n-1} &= -h^2/2\cdot \nabla f(x_{[n-1,n]})-h^2/2 \cdot \nabla f(x_{[n,n+1]})
\end{align*}
where $x_{[n-1,n]}$ denotes a random approximation of $x$ between time $(n-1)h$ and $nh$. The two algorithms of course coincide when the points are deterministically queried at $x_n$, but it is somewhat intuitive that the extra degrees of freedom brought by $x_{[n-1,n]}$ and $x_{[n,n+1]}$ can yield a discretized algorithm of higher fidelity.




\section{Discussion}
We have uncovered several complementary ideas useful for achieving faster rate on quadratic potentials for HMC. In fact, when viewed as ways of reducing the dissipation to avoid random-walk-like behavior, they seem to be different sides of the same coin. Much remains to be understood for the general case, as to how much the same principle can be extended, although the discussion in Section \ref{sec:general} hopefully offers a glimpse of hope. The various approaches we studied all seem to hint at the same $\mathcal{O}(\frac{1}{\sqrt{\mu}}\log(1/\epsilon))$ rate, which also provides an impetus to investigate the tightness of the rate for methods based on Hamiltonian dynamics generally. In a different vein, exploiting randomness for designing integrators for Hamiltonian ODE deserves better treatment and seems largely missing from the classical literature \cite{leimkuhler2004simulating}.

\bibliographystyle{plain}
\bibliography{main}

\begin{thebibliography}{10}

\bibitem{bou2021hmc-mixing}
Nawaf Bou-Rabee and Andreas Eberle.
\newblock {Mixing time guarantees for Unadjusted Hamiltonian Monte Carlo}.
\newblock {\em arXiv preprint arXiv:2105.00887}, 2021.

\bibitem{bou2022unadjusted}
Nawaf Bou-Rabee and Milo Marsden.
\newblock {Unadjusted Hamiltonian MCMC with Stratified Monte Carlo Time
  Integration}.
\newblock {\em arXiv preprint arXiv:2211.11003}, 2022.

\bibitem{randomized_hmc}
Nawaf Bou-Rabee and Jes{\'u}s~Mar{\'\i}a Sanz-Serna.
\newblock {Randomized Hamiltonian Monte Carlo}.
\newblock {\em The Annals of Applied Probability}, 27(4):2159--2194, 2017.

\bibitem{carpenter2017stan}
Bob Carpenter, Andrew Gelman, Matthew~D Hoffman, Daniel Lee, Ben Goodrich,
  Michael Betancourt, Marcus~A Brubaker, Jiqiang Guo, Peter Li, and Allen
  Riddell.
\newblock {Stan: A probabilistic programming language}.
\newblock {\em Journal of statistical software}, 76, 2017.

\bibitem{chen_vempala2022optimal}
Zongchen Chen and Santosh~S Vempala.
\newblock {Optimal Convergence Rate of Hamiltonian Monte Carlo for Strongly
  Logconcave Distributions}.
\newblock {\em Theory of Computing}, 18(1):1--18, 2022.

\bibitem{dalalyan2020sampling}
Arnak~S Dalalyan and Lionel Riou-Durand.
\newblock {On sampling from a log-concave density using kinetic Langevin
  diffusions}.
\newblock {\em Bernoulli}, 26(3):1956--1988, 2020.

\bibitem{PDMP}
George Deligiannidis, Daniel Paulin, Alexandre Bouchard-C{\^o}t{\'e}, and
  Arnaud Doucet.
\newblock {Randomized Hamiltonian Monte Carlo as scaling limit of the bouncy
  particle sampler and dimension-free convergence rates}.
\newblock {\em The Annals of Applied Probability}, 31(6):2612--2662, 2021.

\bibitem{duane1987hybrid}
Simon Duane, Anthony~D Kennedy, Brian~J Pendleton, and Duncan Roweth.
\newblock {Hybrid monte carlo}.
\newblock {\em Physics letters B}, 195(2):216--222, 1987.

\bibitem{u-turn}
Matthew~D Hoffman, Andrew Gelman, et~al.
\newblock {The No-U-Turn sampler: adaptively setting path lengths in
  Hamiltonian Monte Carlo.}
\newblock {\em J. Mach. Learn. Res.}, 15(1):1593--1623, 2014.

\bibitem{horowitz1991generalized}
Alan~M Horowitz.
\newblock {A generalized guided Monte Carlo algorithm}.
\newblock {\em Physics Letters B}, 268(2):247--252, 1991.

\bibitem{leimkuhler2004simulating}
Benedict Leimkuhler and Sebastian Reich.
\newblock {\em {Simulating Hamiltonian Dynamics}}.
\newblock Number~14. Cambridge university press, 2004.

\bibitem{lu2020explicit}
Jianfeng Lu and Lihan Wang.
\newblock {On explicit L2-convergence rate estimate for piecewise deterministic
  Markov processes}.
\newblock {\em arXiv preprint arXiv:2007.14927}, 2020.

\bibitem{monmarche2022hmc}
Pierre Monmarch{\'e}.
\newblock {HMC and Langevin united in the unadjusted and convex case}.
\newblock {\em arXiv preprint arXiv:2202.00977}, 2022.

\bibitem{neal2011mcmc}
Radford~M Neal et~al.
\newblock {MCMC using Hamiltonian dynamics}.
\newblock {\em Handbook of markov chain monte carlo}, 2(11):2, 2011.

\bibitem{malt2022}
Lionel Riou-Durand and Jure Vogrinc.
\newblock {Metropolis Adjusted Langevin Trajectories: a robust alternative to
  Hamiltonian Monte Carlo}.
\newblock {\em arXiv preprint arXiv:2202.13230}, 2022.

\bibitem{vishnoi2021introduction}
Nisheeth~K Vishnoi.
\newblock {An introduction to Hamiltonian Monte Carlo method for sampling}.
\newblock {\em arXiv preprint arXiv:2108.12107}, 2021.

\bibitem{wang2022accelerating}
Jun-Kun Wang and Andre Wibisono.
\newblock {Accelerating Hamiltonian Monte Carlo via Chebyshev Integration
  Time}.
\newblock {\em arXiv preprint arXiv:2207.02189}, 2022.

\end{thebibliography}

\end{document}